# Generation of $^{87}$Rb-resonant bright two-mode squeezed light with four-wave mixing

SAESUN KIM AND ALBERTO M. MARINO*

*Homer L. Dodge Department of Physics and Astronomy, The University of Oklahoma, Norman, Oklahoma 73019, USA*
*marino@ou.edu

**Abstract:** Squeezed states of light have found their way into a number of applications in quantum-enhanced metrology due to their reduced noise properties. In order to extend such an enhancement to metrology experiments based on atomic ensembles, an efficient light-atom interaction is required. Thus, there is a particular interest in generating narrow-band squeezed light that is on atomic resonance. This will make it possible not only to enhance the sensitivity of atomic based sensors, but also to deterministically entangle two distant atomic ensembles. We generate bright two-mode squeezed states of light, or twin beams, with a non-degenerate four-wave mixing (FWM) process in hot $^{85}$Rb in a double-lambda configuration. Given the proximity of the energy levels in the D1 line of $^{85}$Rb and $^{87}$Rb, we are able to operate the FWM in $^{85}$Rb in a regime that generates two-mode squeezed states in which both modes are simultaneously on resonance with transitions in the D1 line of $^{87}$Rb, one mode with the $F = 2$ to $F' = 2$ transition and the other one with the $F = 1$ to $F' = 1$ transition. For this configuration, we obtain an intensity difference squeezing level of $-3.5$ dB. Moreover, the intensity difference squeezing increases to $-5.4$ dB and $-5.0$ dB when only one of the modes of the squeezed state is resonant with the D1 $F = 2$ to $F' = 2$ or $F = 1$ to $F' = 1$ transition of $^{87}$Rb, respectively.

## 1. Introduction

Squeezed states of light have gained a significant amount of attention given that they exhibit noise levels below the quantum noise limit (QNL). This property makes it possible to enhance the sensitivity of sensors [1–3], the security of protocols for quantum key distribution [4–6], and the resolution of image processing [7, 8]. More recently, squeezed states of light have also been proposed as a way to realize one-way quantum computing [9, 10].

The ability to enhance the sensitivity of a device with squeezed states can be extended to systems based on atomic ensembles. In order to achieve this, the light needs to be near or on atomic resonance to obtain an efficient interaction with the atomic ensemble. This is particularly true for systems based on cold atoms for which the linewidth of the transition is determined by the natural linewidth of the atom and is of the order of 10 MHz. Among other things, realizing an efficient interaction between an atomic ensemble and squeezed light will make it possible to enhance atomic sensors. For example, previous studies proposed and demonstrated that resonant squeezed light can enhance the sensitivity of atomic interferometers [11, 12], atomic magnetometers [13, 14], and atomic clocks [15]. Moreover, resonant squeezed light makes it possible to study the interaction of non-classical light with atomic ensembles. For example, it has been shown that electromagnetically induced transparency (EIT) preserves the quantum properties of squeezed light [16, 17] and that it can be used to implement a quantum memory for this quantum state of light [18, 19].

Squeezed light can be generated with a number of different systems, such as non-linear crystals [20–22], optical fibers [23], semi-conductor lasers [24], and atomic systems [25, 26]. However, efficient generation of narrow-band squeezed light on resonance with an atomic transitions has proven to be a challenge. For instance, its generation with an optical parametric oscillator (OPO) requires pumping the nonlinear crystal with UV light. At these frequencies, the parametric-down conversion process is not efficient for the generation of squeezed light [27, 28]. In addition, crystal-based sources tend to be complicated and require special manufacturing of the nonlinear crystal. On the other hand, while atomic-based sources should offer a natural and more direct way to generate narrow-band on-resonance quantum states of light, the strong absorption associated with working on resonance is a significant limiting factor.

Despite these complications, narrow-band on-resonance squeezed light has been generated. For instance, single-mode vacuum squeezed light resonant with the Rb D1 line has been generated with a periodically poled KTP crystal. With this system $-4$ dB of squeezing with bandwidths of 150 kHz to 500 kHz was initially obtained [27] and recently improved to $-5.6$ dB of squeezing [29]. Similar systems have also been used to generate $-3$ dB of vacuum single-mode squeezed light on resonance with the D2 line of Cs [30]. In terms of atomic-based sources, polarization self-rotation in a vapor cell [31] has been used to generate vacuum single-mode squeezing on resonance with the D1 ($-3$ dB of squeezing) [32] and D2 ($-1$ dB of squeezing) lines of Rb [33].

As opposed to previous work that has focused on the generation of single-mode squeezing on resonance with atomic transitions, we concentrate on the generation of on-resonance two-mode squeezed light. For this purpose, four-wave mixing (FWM) in a double-lambda configuration is a good candidate as it has been shown to produce $-9$ dB of intensity difference squeezing when operating off resonance [34–36]. In recent years, there has been a significant interest on the generation of two-mode squeezed states due to their entanglement properties, as this makes them the standard source for continuous-variable (CV) entangled states and CV quantum information [37]. In particular, the ability to generate these quantum states of light on resonance with an atomic transition will make it possible to deterministically transfer the entanglement from the light to two remote atomic ensembles and to generate entangled atom laser beams [38].

In this work, we generate bright two-mode squeezed states of light on resonance with the D1 line of $^{87}$Rb with a FWM process in a $^{85}$Rb vapor cell. To do this, we take advantage of the proximity of the D1 energy levels of the two rubidium isotopes. As a result, we are able to operate the FWM process in a configuration in which one mode can be on resonance with the D1 $F = 2$ to $F' = 2$ transition while the other mode can be on resonance with the D1 $F = 1$ to $F' = 1$ transition of $^{87}$Rb. For a configuration in which both modes are on resonance, we observe −3.5 dB of intensity-different squeezing. On the other hand, we observe −5.4 dB and −5.0 dB of squeezing when only one of the modes of the squeezed state is on resonance with the D1 $F = 2$ to $F' = 2$ or $F = 1$ to $F' = 1$ transition, respectively.

## 2. Experimental setup

We generate bright two-mode squeezed states of light using a non-degenerate FWM process in a double-lambda configuration in the D1 line of $^{85}$Rb, as shown in Fig. 1. In this process, two photons from a single pump beam are converted into a pair of photons, which we call probe and conjugate. This leads to the amplification of the input probe beam and the generation of the conjugate. The experimental setup is shown in Fig. 2. We use a CW Ti:sapphire laser to generate the pump beam required for the FWM. In order to generate the probe beam, we pick-off a small portion of the pump before the Rb cell with a beam sampler and double pass this beam through an acoustic optical modulator (AOM) to generate the required frequency for the probe. For an efficient FWM process, the frequency difference between the pump and the probe needs to be close to the separation between the two ground state hyperfine levels of $^{85}$Rb, such that the pump and probe are on two-photon resonance with the corresponding Raman transition. Both the pump and probe are sent through optical fibers to clean up their spatial modes. After the fibers, perpendicularly polarized pump and probe beams intersect at an angle $\theta$ at the center of a 12 mm $^{85}$Rb vapor cell with anti-reflection coated windows. After the FWM process, the pump beam is blocked with a polarization filter and the probe and conjugate beams are measured with a balanced detection system with 95% quantum efficient photodiodes.

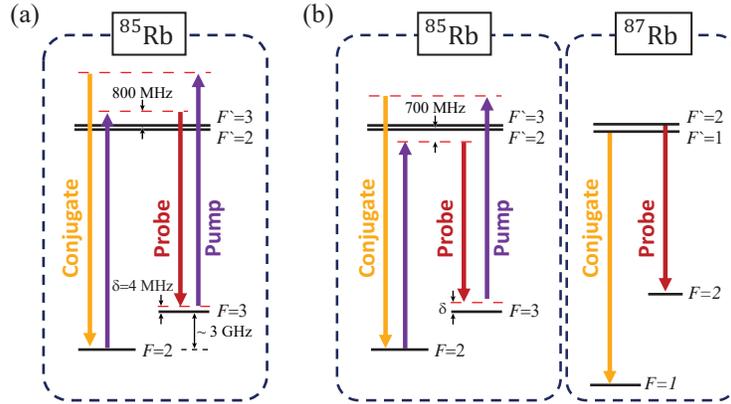

Fig. 1. Energy level diagram of the double-lambda system in the D1 line of $^{85}$Rb. (a) Optimized FWM configuration for the generation of off-resonance two-mode squeezed state with −9 dB of squeezing. (b) When the one-photon detuning of the pump is tuned ∼ 700 MHz to the red of the D1 $F = 2$ to $F'$ transition of $^{85}$Rb (figure on left), the generated probe and conjugate can be resonant with the D1 $F = 2$ to $F' = 2$ and $F = 1$ to $F' = 1$ transitions in $^{87}$Rb (figure on right).

This process has been previously shown to generate −9 dB of off-resonance intensity difference squeezing [34, 36] as well as large levels of CV entanglement [7]. In this configuration, the probe and conjugate beams are around 800 MHz and 3.8 GHz to the blue of their corresponding transition in the D1 line of $^{85}$Rb, respectively, as shown in Fig. 1(a). The typical parameters for the off-resonance configuration are a $1/e^2$ diameter for the pump and the probe of 1.3 mm and 0.7 mm, respectively, angle $\theta$ of 0.3°, pump power of ≈400 mW, and a two-photon detuning $\delta$ of 4 MHz [34]. While in principle the optimum FMW process should occur at $\delta = 0$, a slight detuning is needed to compensate for the light shift introduced by the strong pump beam

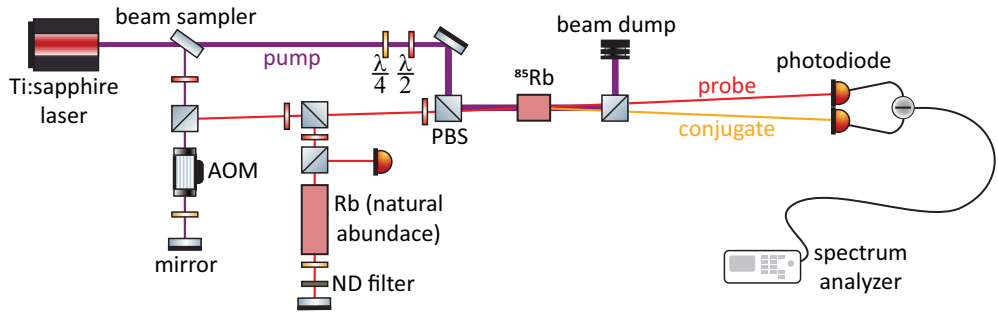

Fig. 2. Simplified schematic of the experimental setup. The pump and probe beams intersect at a slight angle inside a $^{85}$Rb vapor cell. As a result of the FWM, the probe beam is amplified and a new beam, the conjugate, is generated. After the cell, the pump beam is filtered with a PBS, and the probe and conjugate intensity difference noise is measured with a spectrum analyzer. A saturated absorption spectroscopy setup is used to obtain a measure of the absolute frequency of the probe beam. AOM: acousto-optic modulator, ND filter: neutral density filter, PBS: polarizing beam splitter.

coupling the D1 $F = 3$ to $F'$ transition of $^{85}$Rb [39].

While the original proposals for the double-lambda configuration [40–42] suggest that it should be possible to use EIT to eliminate absorption when operating on resonance, in practice this is not the case and residual absorption dominates over the FWM process when the probe frequency is tuned closer to resonance. In order to overcome this limitation and obtain on-resonance squeezed states of light, we take advantage of the proximity of the energy levels of the D1 line in $^{85}$Rb and $^{87}$Rb, as shown in Fig. 1(b). In particular, the frequency difference between the fine structure for the D1 transitions of the Rb isotopes is relatively small ($\approx$78 MHz), while the frequency difference between the hyperfine transitions with the largest energy separation in each isotope ($F = 2$ to $F' = 3$ in $^{85}$Rb and $F = 1$ to $F' = 2$ in $^{87}$Rb) is 3.8 GHz. This makes it possible to tune the pump and probe for the FWM in $^{85}$Rb such that either or both of the twin beams are resonant with a transition in the D1 line of $^{87}$Rb. In particular, by detuning the pump by around 700 MHz to the red of the $F = 2$ to $F'$ transition in $^{85}$Rb, as shown in Fig. 1(b), we are able to bring the probe and/or conjugate to resonance with the $F = 2$ to $F' = 2$ and $F = 1$ to $F' = 1$ transitions in the D1 line of $^{87}$Rb, respectively.

We consider different configurations for the FWM in which either the probe, conjugate, or both beams are on exact resonance. For each configuration, in order to obtain an absolute measure of the frequency of the probe, we take a portion of the probe beam before the Rb vapor cell and send it to a saturation spectroscopy setup, as shown in Fig. 2. This allows us to precisely determine the transition to which the probe beam is tuned to. Finally, to determine the absolute frequency of the conjugate, we measure the beat note between the probe and conjugate beams after the FWM process.

## 2.1. Singly-resonant configuration

We first consider the configuration in which only the probe or the conjugate is on resonance with one of the transitions of $^{87}$Rb. In this configuration the one-photon detuning is determined by the frequency of the $^{87}$Rb transition; however, the two-photon detuning can be optimized to obtain the maximum level of squeezing. For this configuration, the optimum level of squeezing is obtained for an angle $\theta$ of 0.45°, pump power and $1/e^2$ diameter of 1 W and 1.9 mm, respectively, $1/e^2$ probe diameter of 0.6 mm, temperature of the vapor cell of 89°C, and two-photon detuning $\delta$ of 16 MHz.

As shown in Fig. 3(a), we are able to tune the frequency of the probe or the conjugate to be on resonance by changing the frequency of the laser. This allows us to tune the one-photon detuning while keeping the two-photon detuning fixed. With a one-photon detuning for the pump of 687 MHz to the red of the D1 $F = 2$ to $F' = 2$ transition of $^{85}$Rb the probe becomes resonant with the D1 $F = 2$ to $F' = 2$ transition of $^{87}$Rb with a measured level of intensity difference squeezing of −5.4 dB at an analysis frequency of 750 kHz. While for a one-photon detuning of 771 MHz, the conjugate becomes resonant with the D1 $F = 1$ to $F' = 1$ transition of $^{87}$Rb with a measured level of squeezing of −5.0 dB at an analysis frequency of 750 kHz. After subtracting the electronic noise, the squeezing level for the case of resonant probe or resonant conjugate is increased to −6.3 dB and −6.2 dB, respectively. Figure 3(b) shows the measured level of squeezing as we tune the laser frequency. As can be seen, it is possible to obtain squeezing over a region of the order of 500 MHz around the resonance.

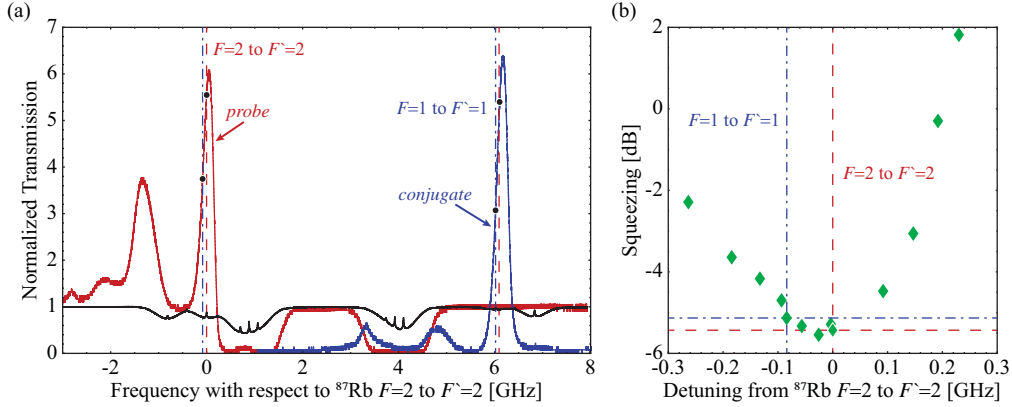

Fig. 3. FWM configuration for resonant probe or conjugate. (a) Transmission spectra for the probe (red trace) and the conjugate (blue trace) generated with FWM in the D1 line of $^{85}$Rb as the one-photon detuning of the pump is scanned while the two-photon detuning is kept fixed at $\delta = 16$ MHz. The black trace shows the saturated absorption spectroscopy spectrum for a natural abundance Rb cell and serve as a frequency reference. The vertical red dashed (blue dashed-dotted) lines indicate the frequencies for the probe and the conjugate for the configuration in which the probe (conjugate) is on resonance with the D1 $F = 2$ to $F' = 2$ ($F = 1$ to $F' = 1$) transition of $^{87}$Rb. The black dots indicate the normalized transmission for the probe and the conjugate for each of these frequencies. (b) Intensity difference squeezing as a function of the detuning of the probe from the D1 $F = 2$ to $F' = 2$ transition in $^{87}$Rb. Squeezing is present over a range of 500 MHz around the resonance frequency. The vertical and horizontal red dashed (blue dashed-dotted) lines indicate the frequency at which the probe (conjugate) is on resonance and the corresponding level of squeezing, respectively. Spectrum analyzer set to a resolution bandwidth of 30 kHz and a video bandwidth of 100 Hz.

The measured levels of squeezing represent a significant decrease from the −9 dB that have been obtained with this system for the off-resonance configuration. This is a result of having the one-photon detuning determined by the frequency of the transition. As a result, the conjugate field is now closer to resonance for the D1 $F = 2$ to $F'$ transition of $^{85}$Rb. This leads to additional residual absorption due to the atomic system, which limits the levels of squeezing that can be achieved. In addition, the constraint on the one-photon detuning limits our ability to optimize the FWM process while minimizing other competing nonlinear processes in the system.

For the optimized parameters, the beat note between the probe and the conjugate, which is determined by the ground state hyperfine splitting of $^{85}$Rb and the two-photon detuning, is measured to be 6.104 GHz. This is consistent with a two-photon detuning $\delta$ of 16 MHz. When compared with the optimum off-resonance configuration, $\delta$ is larger by a factor of 4. We can understand this by noting that the detuning of the pump coupling the D1 $F = 3$ to $F'$ transition of $^{85}$Rb is reduced from ~4 GHz to ~2 GHz, while the pump power required for the FWM is increased from ~500 mW to 1 W. This leads to an increase in the light shift by a factor of 4, consistent with the increase for the two-photon detuning. Thus effectively, the system is kept on two-photon resonance for the light shifted Raman transition.

With this technique it is not possible to generate quantum states of light resonant with the D1 $F = 1$ to $F' = 2$ or $F = 2$ to $F' = 1$ transition of $^{87}$Rb. The generation of squeezed light on resonance with the $F = 1$ to $F' = 2$ transition would require a one-photon detuning for the pump of 42 MHz to the blue of the $F = 2$ to $F' = 2$ transition in $^{85}$Rb. This leads to a probe frequency too close to the $F = 3$ to $F'$ transition of $^{85}$Rb, which results in absorption dominating over the FWM process. On the other hand, the generation of squeezed light on resonance with $F = 2$ to $F' = 1$ is also challenging. This case requires a "symmetric" configuration in which one of the Raman transitions of the double-lambda configuration is detuned to the red of the D1 transition in $^{85}$Rb by 1.86 GHz while the other is detuned to the blue by 1.17 GHz. Under these conditions, the response from the two lambda systems interferes destructively and no gain is obtained with the FWM process [43].

## 2.2. Doubly-resonant configuration

Next, we explore the possibility of having both the probe and conjugate simultaneously on resonance with the D1 $F = 2$ to $F' = 2$ and $F = 1$ to $F' = 1$ transitions in $^{87}$Rb, respectively. In addition to fixing the one-photon detuning of the pump to 687 MHz to the red of the D1 $F = 2$ to $F' = 2$ transition of $^{85}$Rb, this requires the frequency difference between the probe and the conjugate to be 6.02 GHz. Such a configuration is possible by setting the two-photon detuning $\delta$ to $-26$ MHz, which corresponds to a frequency difference between the probe and pump of 3.01 GHz. Given these constraints, the optimum level of squeezing is obtained for an angle $\theta$ of 0.5°, pump power and $1/e^2$ diameter of 1 W and 1.9 mm, respectively, $1/e^2$ probe diameter of 0.6 mm, and temperature of the vapor cell of 91°C.

Figure 4(a) shows the transmission spectra for the probe and the conjugate for the doubly-resonant configuration. For the optimized parameters, we are able to directly measure $-3.5$ dB of intensity difference squeezing at an analysis frequency of 750 kHz when both probe and conjugate are simultaneously on resonance. The level of squeezing increases to $-3.9$ dB after subtraction of the electronic noise. To verify that both modes are simultaneously on resonance, we set the frequency of the probe to the $^{87}$Rb D1 $F = 2$ to $F' = 2$ transition peak of the saturated absorption spectrum, as shown in Fig. 4(a). We then verify the frequency of the conjugate by measuring the beat note between the probe and the conjugate. As expected, the beat-note frequency was measured to be 6.02 GHz.

It is worth noting that squeezing is still present when the laser frequency is tuned over a range of 300 MHz around the resonance, as shown in Fig. 4(b). Both the level of squeezing and the tunabiltiy are significantly reduced with respect to the singly-resonant and off-resonant configurations. This is a result of having to fix the two-photon detuning in addition to the one-photon detuning. In this configuration the FWM is no longer on two-photon resonance, as required for an efficient FWM process.

The ability to generate twin beams of light in which both modes are simultaneously on resonance with an atomic transition opens the door for the deterministic transfer of entanglement from the light to two remote atomic ensembles. Through the implementation of an EIT based quantum memory, which has been shown to preserve the quantum properties of light [18, 19], in two remote atomic ensembles it is possible to simultaneously transfer the properties of the probe and the conjugate to the spin degree of freedom of the memories. This will lead to the deterministic remote entanglement of the two atomic ensembles, as well as the possibility of

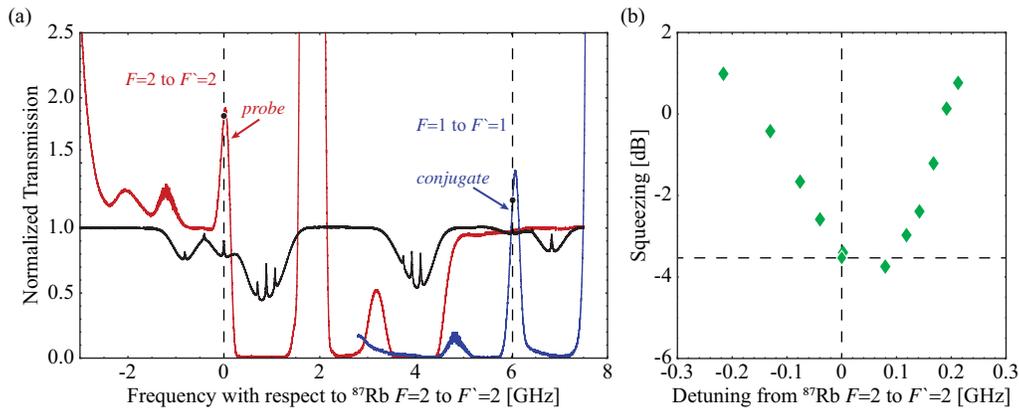

Fig. 4. FWM configuration for probe and conjugate simultaneously on resonance. (a) Transmission spectra for the probe (red trace) and the conjugate (blue trace) generated with FWM in the D1 line of $^{85}$Rb as the one-photon detuning of the pump is scanned while the two-photon detuning is kept fixed at $\delta = -26$ MHz. The black trace shows the saturated absorption spectroscopy spectrum for a natural abundance Rb cell and serve as a frequency reference. The vertical dashed lines indicate the resonance frequencies for the probe and the conjugate with the D1 $F = 2$ to $F' = 2$ and $F = 1$ to $F' = 1$ transitions of $^{87}$Rb, respectively. The black dots indicate the normalized transmission for the probe and the conjugate for each of these frequencies. (b) Intensity difference squeezing as a function of the detuning of the probe from the D1 $F = 2$ to $F' = 2$ transition in $^{87}$Rb. Squeezing is present over a range of 300 MHz around the resonance frequency. The vertical and horizontal dashed lines indicate the resonance frequency for the probe and conjugate and the corresponding level of squeezing, respectively. Spectrum analyzer set to a resolution bandwidth of 30 kHz and a video bandwidth of 100 Hz

storage and delay of CV entangled beams of light.

## 3. Conclusion

We have demonstrated the generation of bright two-mode squeezed light in which the probe and/or the conjugate can be on resonance with different transitions of $^{87}$Rb using a non-degenerate FWM process in the D1 line of $^{85}$Rb. For the doubly-resonance configuration in which the probe and conjugate are simultaneously on resonance, we obtain $-3.5$ dB of intensity difference squeezing. The intensity difference squeezing increases to $-5.4$ dB and $-5.0$ dB when only one of the beams is tuned to resonance. The generation of squeezed light using atomic ensembles overcomes several experimental difficulties associated with crystal-based sources. In addition, resonant squeezed states with bandwidths of the order of the atomic natural linewidth make it possible to study the interaction between atoms and quantum states of light. Among other things, it will make it possible to enhance quantum devices based on atomic systems and to deterministically entangle two remote atomic ensembles.

## Funding

Air Force Office of Scientific Research (AFOSR) (FA9550-15-1-0402).

## Acknowledgments

The authors would like to thank Gaurav Nirala and Tim Woodworth for constructive criticism of the manuscript and Ashok Kumar for insightful conversations.